\documentclass[namedreferences]{solarphysics}

\usepackage{amssymb, amsmath, amsfonts}
\usepackage[hyperref,optionalrh]{spr-sola-addons} % For Solar Physics 
\usepackage[optionalrh]{spr-sola-addons} % For Solar Physics 
\usepackage{graphicx}        % For eps figures, newer & more powerfull
\usepackage{courier}         % Change the \texttt command to courier style
\usepackage{amssymb}        % useful mathematical symbols
\usepackage{color}           % For color text: \color command
\usepackage{breakurl}        % For breaking URLs easily trough lines
\usepackage{float, times}
\usepackage{tabularx}
\usepackage[nice]{nicefrac}
\usepackage[T1]{fontenc}
\renewcommand{\vec}[1]{{\mathbfit #1}}

\newcommand{\aap}{    {\it Astron. Astrophys.}}

\newcommand{\apj}{    {\it Astrophys. J.}}
\newcommand{\apjl}{   {\it Astrophys. J. Lett.}}

\newcommand{\mnras}{  {\it Mon. Not. Roy. Astron. Soc.}}

\newcommand{\solphys}{{\it Solar Phys.}}

\chardef\us=`\_

\begin{document}

	%			\title{Article Preparation Guidelines for {\it Solar Physics}}
	%			
	%			\author[addressref={aff1,aff2,aff3},email={e-mail.a@mail.com}]{\inits{F.N.}\fnm{First~Names}~\lnm{Last~Name~Author-a}\orcid{123-456-7890}}
	%			\author[addressref=aff1,email={e-mail.b@mail.com}]{\inits{F.}\fnm{First~Names}~\lnm{Last~Name~Author-b}}
	%			\author[addressref=aff2,corref,email={e-mail.c@mail.com}]{\inits{F.}\fnm{First~Names}~\lnm{Last~Name~Author-c}\orcid{987-654-3210}}
	%			\author[addressref=aff3]{\inits{T.}\fnm{First~Names}~\lnm{Last~Name~Author-d}}
	%			%\author{\inits{}\fnm{}~\lnm{}\orcid{}}
	%			%   NOTE:  Just one corresponding author [corref]
	%			%   \institute{$^{1}$ First affiliation
		%			%                     email: \url{e.mail-a} email: \url{e.mail-b}\\ 
		%			%              $^{2}$ Second affiliation
		%			%                     email: \url{e.mail-c} \\
		%			%             \textit{}
		%			\address[id=aff1]{First very very very very very very very very very
			%				very very very very very very very very very very very very very very very very very very long affiliation and address}
		%			\address[id=aff2]{Institution, City, State, Country}
		%			\address[id=aff3]{Third affiliation and address}
		%			
		%			\runningauthor{Author-a et al.}
		%			\runningtitle{\textit{Solar Physics} Example Article}
		
\begin{article}			
\begin{opening}

\title{On Strengthening of the Solar f-mode Prior to Active Region Emergence Using the Fourier-Hankel Analysis}
\author[addressref={aff1,aff2},corref,email={mwaidele@stanford.edu}]{Waidele, M.\orcid{0000-0003-2678-626X}}
\author[addressref={aff1,aff3}]{Roth, M.\orcid{0000-0002-1430-7172}}
\author[addressref={aff4}]{Singh, N.K.\orcid{0000-0001-6097-688X}}
\author[addressref={aff5}]{K\"apyl\"a, P.J.\orcid{0000-0001-9619-0053}}
\address[id=aff1]{
	Leibniz-Institut f\"ur Sonnenphysik (KIS), Sch\"oneckstrasse 6, 79104, Germany
}
\address[id=aff2]{
	W. W. Hansen Experimental Physics Laboratory, Stanford University, Stanford, CA 94305-4085,
	USA	
}
\address[id=aff3]{
	Thüringer Landessternwarte, Sternwarte 5, 
	07778 Tautenburg, Germany
}
\address[id=aff4]{
	Inter-University Centre for Astronomy \& Astrophysics, Post Bag 4, Ganeshkhind, Pune 411 007, India
}
\address[id=aff5]{
	Georg-August-Universit\"at G\"ottingen, Institut f\"ur Astrophysik und Geophysik, Friedrich-Hund-Platz 1, 37077 G\"ottingen.
}

\begin{abstract}
Recent results of \cite{2016ApJ...832..120S} show that the emergence of an active region (AR) can be seen in a strengthening of the f-mode power up to two days prior of the region's formation. 
In the original work, ring diagram analysis was used to estimate the power 
evolution. In this study, we make use of the Fourier-Hankel method, essentially testing 
the aforementioned results with an independent method.
The data is acquired from SDO/HMI, studying the ARs 11158, 11072, 11105, 11130, 11242 and 11768. 
Investigating the total power as a function of time, we find a similar behavior to the original work for four of the six listed ARs, which is an enhancement of f-mode power about one to three days prior to AR emergence. AR 11105 as well as AR 11242 also show detectable f-mode enhancement, although less significant than what was observed before.
In addition, the analysis of the absorption coefficient $\alpha$, yielded by the Fourier-Hankel analysis, shows neither absorption ($\alpha > 0$) nor emission ($\alpha < 0$) of power during the enhancement. 
Finding no changes of the absorption coefficient (i.e. $\alpha = 0$) is an important result, as it narrows down the possible physical interpretation of the original f-mode power enhancement, showing that no directional dependence (in the sense of inward and outward moving waves) is present.
%\textcolor{green}{Further analysis including additional active regions is needed for improved statistics.}
\end{abstract}
\keywords{Sun, helioseismology; Sun, activity; Sun, magnetic fields}
\end{opening}

\section{Introduction}
\label{sec:introduction}
Finding ways to reliably predict the emergence of active regions is an important application of helioseismology, as it allows for the ability to forecast some aspects of space weather. 
%Solar eruptive phenomena embrace a variety of eruptions, including flares, solar energetic particles, and radio bursts
Active regions act as hosts to almost all solar eruptive phenomena, such as flares, radio bursts and coronal mass ejections \citep{2012LRSP....9....3W, 2021LRSP...18....4T}.
It has been shown in the past that active regions strongly influence helioseismic measurements: 
For example, in \citet{2001ApJ...563..410R} a significant shift towards larger frequencies for both pressure and fundamental (p- and f-) modes was found, which is related to strong p-mode travel-time perturbations \citep{2006ApJ...647L.187B}. Similarly, \citet{2013AaA...558A.130S} reported a p- and f-mode travel time shift of up to $100\,$s (in a simulated scenario). Both \citet{2013SoPh..287..107R} and \citet{2015ApJ...801...27R} investigate a local enhancement of acoustic power that surrounds all magnetic features on the Sun (the acoustic halo). 
Measurements of subsurface wave-speed perturbation show that the average acoustic power at the surface is altered even before the appearance of a strong magnetic field in surface magnetograms \citep{2011SoPh..268..321H}.
%many solar properties measured with helioseismic methods \citep{2001ApJ...563..410R, 2006ApJ...647L.187B, 2013AaA...558A.130S, 2015ApJ...801...27R}, even before the appearance of a strong magnetic field in surface magnetograms \citep{2009ASPC..416..147H}. 
Emerging active regions have been investigated in the past by means of time-distance helioseismology \citep{2011Sci...333..993I}, down to a depth of 75 Mm. 
In terms of actual precursor signals, a retrograde flow with an amplitude of $100\,$m$/$s roughly 2 hours prior to emergence has been observed \citep{2016AaA...595A.107S}, while subsurface flows remain weak (on the order of $1\,$m$/$s, not larger than $15\,$m$/$s) and obfuscated \citep{2008ApJ...680.1457H, 2013ApJ...762..131B}.
Other studies involving the f-mode yielded results regarding shallow (usually a few Mm) subsurface magnetic fields \citep{2004ApJ...613.1253H}, generally indicating flows that are not present in magnetically quiet regions. 
The theoretical work of \cite{2014ApJ...795L...8S, 2020GApFD.114..196S} shows that f-mode frequencies, as well as amplitudes, are sensitive to subsurface magnetic fields. 
As many details about the build up and the emergence of active regions, as well as their subsurface flows and structure remain uncertain \citep{2009SSRv..144..249G, 2010SoPh..267....1M}, studying f-mode variability in the presence of active regions is a promising method to learn about the solar dynamo \citep{2020LRSP...17....4C} and subsurface solar magnetism more specifically \citep{2008ApJ...680.1457H, 2012ApJ...757..148F}.
	
This study builds on the observations of \cite{2016ApJ...832..120S}, who reported a systematic strengthening of f-mode power, prior to the emergence of active regions ARs 11130, 11158, 11242, 11105, 11072, and 11768.
Here, we employ a Fourier-Hankel analysis on the same ARs to first estimate the power evolution aiming to test the aforementioned observations with an independent method, and secondly, to calculate the absorption coefficient $\alpha$, i.e. the ratio of power from the waves going towards and away from the active region, to potentially link the results to physical mechanisms (such as sunspot absorption), uncovered in earlier studies. 
Fourier-Hankel analyses were first applied in the context of helioseismology by \cite{1987ApJ...319L..27B}, in order to investigate the mentioned power absorption by sunspots. 
The Fourier-Hankel method was further used and refined in later studies, for example \citet{2013SoPh..282...15C} who looked at the absorption spectrum as a whole for multiple HMI and AIA data products, and  \citet{2014ApJ...788..136F}, who investigated the capability of helioseismic methods to discern between monolithic and spaghetti sunspot models. 
Furthermore, it has been useful specifically in investigations of power evolution in relation to solar surface magnetism \citep{1995ApJ...451..877F, 1999ApJ...510..494L}.
	
Our work relies on Dopplergrams recorded by HMI \citep{2012SoPh..275..229S} and
the data processing reported in the original work of \cite{2016ApJ...832..120S}. 
We test the hypothesis that there is an initial enhancement of f-mode power 1 to 3 days prior to the emergence of an active region, followed by the well-established power reduction due to the sunspot being present. 
Simultaneous to the reduction of power, an increase in absorption ($\alpha > 0$) would follow. 
Therefore, calculating $\alpha$ during the initial power enhancement should either result in emission (decrease of the absorption coefficient, $\alpha < 0$), which would let us make a connection to the sunspot-power behavior, or a result of zero absorption ($\alpha = 0$) suggesting another, potentially unknown, mechanism to be present.
	
At first we will go over data acquisition and treatment in Section \ref{sec:2}, discussing similarities and differences compared to the original work. 
Second, details regarding the Fourier-Hankel method applied here will be
discussed in Section \ref{sec:3}. 
Our results will be presented in Section \ref{sec:4}, while discussion and conclusions are shown in Section \ref{sec:5}.
	
\section{Data Acquisition and Treatment}
\label{sec:2}
\begin{figure}[htb!]
	\centering
	\includegraphics[width=1.\textwidth]{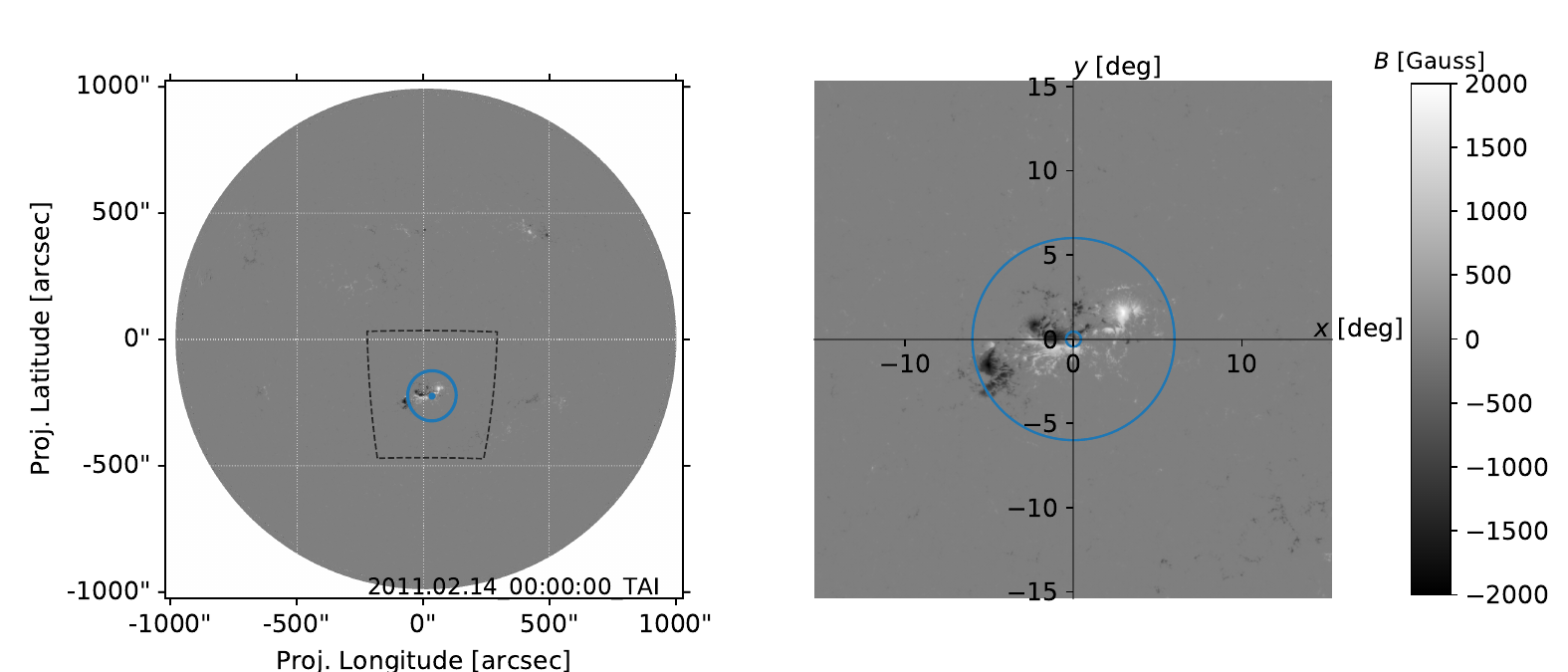}
	\caption{
		Magnetogram of AR 11158 in helioprojective full disk view (left) and zoomed, heliocentric
		view (right) during emergence on 2011.02.13, 00:00, displaying the line-of-sight component of the magnetic field. Black dashed lines mark the full tracked area while blue circles highlight the data that is selected for further analysis. The right panel contains the Postel-projected map from the tracked area marked in the left panel.
	}
	\label{fig:1}
\end{figure}

\begin{figure}[htb!]
	\centering
	\includegraphics[width=1.\textwidth]{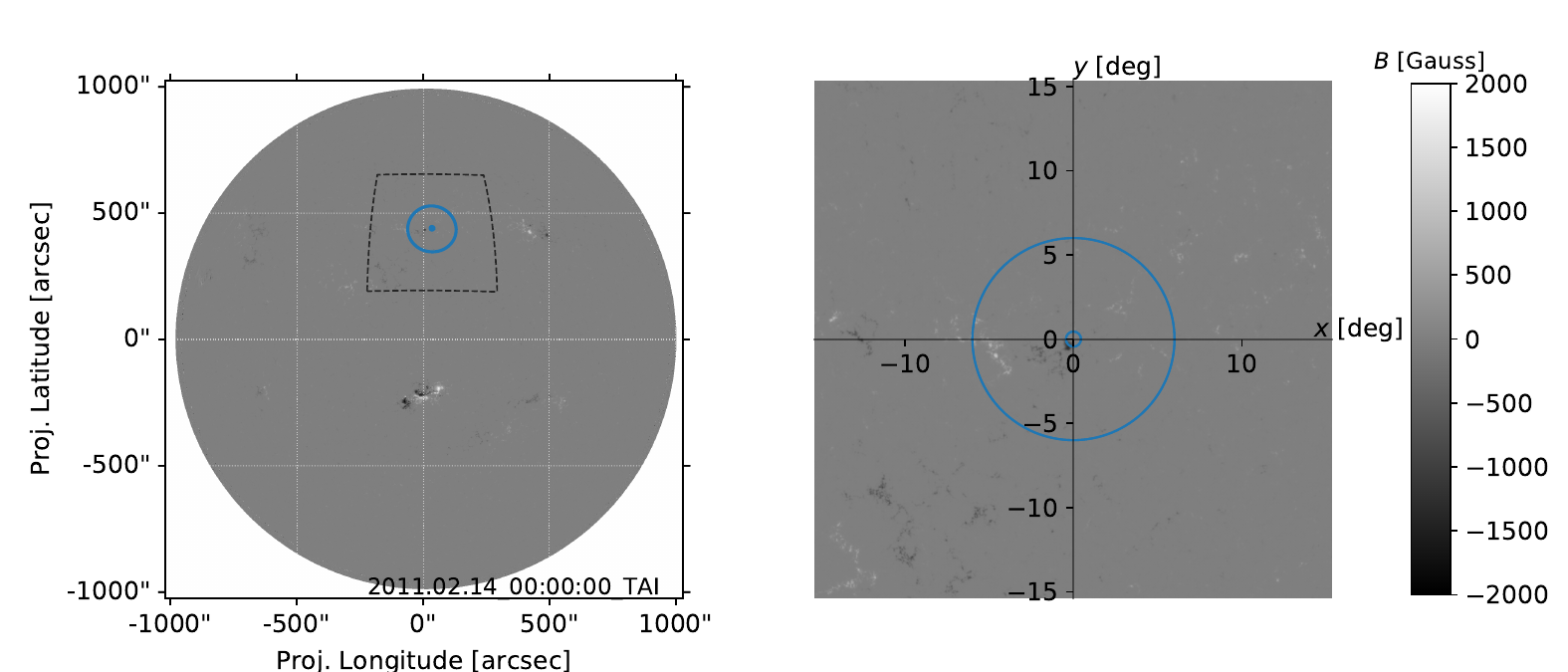}
	\caption{
		Magnetogram of the quiet-Sun region corresponding to AR 11158 (i.e. QS 11158) in helioprojective full disk view (left) and zoomed, heliocentric view (right) during emergence on 2011.02.13, 00:00. As remarked, this region is tracked simultaneously to AR 11158, but on the opposite hemisphere. An apparent offset from the equator ($y=0$) can be seen, which is due to the non zero $B_0$-angle.
	}
	\label{fig:1_2}
\end{figure}

The data preparation is designed to be as similar as possible to that described in \cite{2016ApJ...832..120S}, albeit a different helioseismic method was used.
As a first step for the analysis, we track AR 11158 starting prior to its emergence on 2011.02.14 with the computer program \textit{mtrack} \citep{2007AN....328..352B}, using the Snodgrass-rate as tracking rate \citep{1990ApJ...351..309S}. 
Furthermore, for every frame the line-of-sight component of both observer velocity and solar rotation are removed.
The contributions of granulation to the velocity signal are reduced by applying a running difference, yielding a from now on labeled corrected velocity signal $v$.
In addition to the magnetically active region, a quiet-Sun control region with the same tracking rate and initial position but on the opposite hemisphere is tracked. 
Thus, for each active region that is tracked with starting point $(x, y)$ (representing longitude, latitude), the corresponding quiet-Sun starting point is given by $(x, -y)$.
Let us call this quiet control region as QS 11158, which corresponds to the AR 11158. 
%The $B_0$-angle at that time is $-6.72^\circ$, which affects the measured surface velocity to some degree, although it does so in a uniform manner concerning northern and southern hemisphere.
It is important to keep in mind that the observer's latitudinal position, commonly known as the $B_0$-angle, will introduce a systematic discrepancy between velocities measured in the northern and southern hemisphere. 
As will be demonstrated below, this can be accounted for using geometric considerations.
Nevertheless, we show the $B_0$-angle for all tracked regions in Table \ref{tab:1}.
All other regions (ARs 11072: $t_\text{AR}=$ 2010.05.23, 11105: $t_\text{AR}=$ 2010.08.03, 11130: $t_\text{AR}=$ 2010.11.29, 11242: $t_\text{AR}=$ 2011.06.29 and 11768: $t_\text{AR}=$ 2013.06.14) are tracked, similarly to what is described for AR 11158. % and to the \cite{2016ApJ...832..120S}. 
As an additional control measure to investigate, whether the location of the QS on the opposite hemisphere leads to significantly different results, AR 11242 is tracked using the Carrington-rate and its quiet-Sun region is tracked at the exact same latitude (i.e. at $(x, y)$) but two Carrington rotations prior to the emergence of AR 11242. %, thus serving as a sample with slightly modified input parameters for the tracking and the quiet-Sun control.
We take the same definition of AR emergence time $t_\text{AR}$ as in the original work. 
Each active region is required to be isolated, meaning no significantly strong magnetic field appears in its vicinity during the full tracking period.  
Furthermore, all regions should emerge close to the central meridian, such that the region can be observed up to three days, before and after.
Using data from cycle 24 only, these requirements lead to a naturally small sample size (a total of six in this case). 
Increasing the sample can be achieved by considering non-isolated ARs. 
In this work however, we will restrict ourselves to isolated regions, preferring 'best case' scenarios over a larger amount of data.
For illustration a snapshot of, e.g. AR 11158 and its corresponding quiet Sun control region (QS 11158) are shown in a helioprojective coordinate frame in Figure \ref{fig:1}, and Figure \ref{fig:1_2} respectively. 
Note that we opted to show Magnetograms here for visualization purposes, since the extent of the magnetically active region is easier to see. 
Otherwise Dopplergrams are used throughout the analysis. 
All disk positions for the tracking can be found in Table \ref{tab:1}.
	
\begin{table}[htb!]
	\begin{tabular}{c|c|c|c|c|c|c}
		& AR 11158 & AR 11130 & AR 11768 & AR 11105 & AR 11242 &  AR 11072 \\
		\hline\hline
		$B_0$ & $-6.72^\circ$ & $1.17^\circ$ & $0.90^\circ$ & $5.91^\circ$ & $2.59^\circ$ & $-1.77^\circ$ \\
		\hline
		latitude & $-20.06^\circ$ & $13.28^\circ$ & $-11.57^\circ$ & $18.89^\circ$ & $17.42^\circ$ & $-15.27^\circ$ \\
		\hline
		longitude & $-49.28^\circ$ & $-36.71^\circ$ & $-16.95^\circ$ & $-26.35^\circ$ & $-36.29^\circ$ & $-41.40^\circ$\\
		& $-45.01^\circ$ & $-32.40^\circ$ & $-12.61^\circ$ & $-22.05^\circ$ & $-31.56^\circ$ & $-37.08^\circ$ \\
		& $-40.70^\circ$ & $-28.05^\circ$ & $-8.23^\circ$ & $-17.72^\circ$ & $-26.83^\circ$ & $-32.73^\circ$ \\
		& $-36.41^\circ$ & $-23.70^\circ$ & $-3.85^\circ$ & $-13.41^\circ$ & $-22.11^\circ$ & $-28.37^\circ$ \\
		& $-32.15^\circ$ & $-19.38^\circ$ & $0.49^\circ$ & $-9.12^\circ$ & $-17.38^\circ$ & $-24.05^\circ$ \\
		& $-27.84^\circ$ & $-15.03^\circ$ & $4.87^\circ$ & $-4.79^\circ$ & $-12.65^\circ$ & $-19.70^\circ$ \\
		& $-23.55^\circ$ & $-10.68^\circ$ & $9.24^\circ$ & $-0.48^\circ$ & $-7.92^\circ$ & $-15.34^\circ$ \\
		& $-19.28^\circ$ & $-6.37^\circ$ & $13.58^\circ$ & $3.82^\circ$ & $-3.19^\circ$ & $-11.02^\circ$ \\
		& $-14.97^\circ$ & $-2.02^\circ$ & $17.96^\circ$ & $8.15^\circ$ & $1.53^\circ$ & $-6.67^\circ$ \\
		& $-10.68^\circ$ & $2.33^\circ$ & $22.34^\circ$ & $12.46^\circ$ & $6.26^\circ$ & $-2.31^\circ$ \\
		& $-6.42^\circ$ & $6.64^\circ$ & $26.67^\circ$ & $16.75^\circ$ & $10.99^\circ$ & $2.01^\circ$ \\
		& $-2.11^\circ$ & $11.00^\circ$ & $31.06^\circ$ & $21.08^\circ$ & $15.72^\circ$ & $6.36^\circ$ \\
		& $2.18^\circ$ & $15.34^\circ$ & $35.43^\circ$ & $25.39^\circ$ & $20.44^\circ$ & $10.72^\circ$ \\
		& $6.45^\circ$ & $19.66^\circ$ & $39.77^\circ$ & $29.69^\circ$ & $25.17^\circ$ & $15.04^\circ$ \\
		& $10.76^\circ$ & $24.01^\circ$ & $44.15^\circ$ & $34.02^\circ$ & $29.90^\circ$ & $19.39^\circ$ \\
		& $15.05^\circ$ & $28.36^\circ$ & $48.52^\circ$ & $38.33^\circ$ & $34.63^\circ$ & $23.75^\circ$ \\
	\end{tabular}
	\caption{
		Additional details concerning the tracking procedure of our six active regions. The upper row shows the AR-number, the following contain the $B_0$-angle the (fixed) latitude, and the 16 disk positions in longitude. Note that AR 11242 was tracked using the Carrington-rate, yielding a slightly different spacing in between time-steps compared to all other ARs. 
	}
	\label{tab:1}
\end{table}

Tracking yields five days of 8 hour long 1024 pixel$\times$1024 pixel-data cubes at a cadence of 45s and a map scale of $0.365\,$Mm ($=0.03^\circ$).
Each of these 8 hour cubes represents one data point of f-modal power, as we will see in the further analysis. 
All data cubes contain Postel-projected (see right panels of Figure \ref{fig:1} \& \ref{fig:1_2}) Dopplergrams in a heliographic longitude-latitude grid, i.e. $v(x, y, t)$.
After performing a Fourier-transform, we select data at $k_x = 0$, i.e. $V(k_x=0, k_y, \nu)$. 
Here, $V$ is the Fourier-transform of $v$, $k_x$ represents the longitudinal component of the wavenumber $\vec{k}$, while $k_y$ represents the latitudinal component and finally $\nu$ is the frequency. 
Afterwards, all data within a certain annulus are considered, as shown in Figure \ref{fig:1}. 
The $k_x=0$ selection is made to be in accordance with \cite{2016ApJ...832..120S}. 
Its benefits are first and foremost to reduce the total amount of data and can be understood as a customary procedure, equivalent to a longitudinal average, yielding the thus obtained mean power variation in our analysis.

\section{Fourier-Hankel decomposition}
\label{sec:3}

The decomposition method used here is described in detail in \cite{1987ApJ...319L..27B} and in \cite{2005LRSP....2....6G}. 
After Fourier transforming the velocity amplitudes $v(x, y, t)$ of all Dopplergrams within a given annulus (Figures \ref{fig:1} and \ref{fig:1_2}), we select $k_x = 0$ as mentioned and interpolate the resulting signal onto a radial grid. 
This allows us to further truncate the data, leaving only signals within an annulus, covering the direct vicinity of the (not yet emerged) active region.
The annulus (defined by two circles with radii $r_\text{inner}, r_\text{outer}$) has the dimension $r_\text{inner}, r_\text{outer} = \left( 15\text{ pixel}, 200\text{ pixel} \right) = \left( 5.48, 73.07 \right)$ Mm, yielding a grid-spacing on the axis of the harmonic degree of $\varDelta \ell=59.50$ (with $\ell=k_yR_\odot$). 
The choice of this small annulus was made since we will deal with very small scale oscillations in the analysis ($\ell \in [1200, 2000]$) and thus sensitivity to short wavelengths is required. 
In previous analyses that applied the Fourier-Hankel method, the magnetic region is usually dropped to exclude direct effects of the sunspot to the velocity signal \citep{1987ApJ...319L..27B, 1995ApJ...451..877F}.
Such direct effects include degrading of the Doppler signal by distorting the atomic line and suppressing oscillations, but also contributions to the line-of-sight velocity by the moat-flow or other surface flows.
Here we do not have to worry about direct effects, since we observe the velocity signal prior to the emergence of the active region.
Furthermore, f-mode power strengthening is considered an indirect effect and thus does not suffer from any such effects introduced by choosing a small annulus.
Regardless, it must be kept in mind that any analysis after the emergence of the active regions is using velocity signals corrupted by direct effects introduced by magnetic regions.
The above mentioned interval of $\ell$ is expected to show f-mode strengthening prior to the emergence of an active region \citep{2016ApJ...832..120S}, and has been shown to exhibit power enhancement due to a localized subsurface magnetic concentration in simulations of \cite{2020GApFD.114..196S}. 
%The region itself is normally excluded by the annulus, since strong magnetic field can degrade the Doppler signal by distorting the atomic line and suppressing oscillations. 
%Since we are mostly interested in the Doppler signal before the magnetic field appears at the surface however, our choice of $r_\text{inner}$ is justified. 
%At $r = r_\text{outer}$, the transformed data $V(r, \varphi, t)$ is critically sampled, such that a third order interpolation is required for $V(r < r_\text{outer}, \varphi, t)$.
For spatial truncating, a Hann-window $\lambda_N$ \citep{1958DoverPubl.} was applied to the data $V(r, t)$ with

\begin{align}
\lambda_N(n) = 0.5 - 0.5\cos\left( \frac{2\pi n}{N-1} \right){ ,}
\label{eq:w}
\end{align}

\noindent where $N$ is the sample size ($N = r_\text{inner} - r_\text{outer} = 185$ pixels for the spatial window and $N = 640$ pixels for the temporal window) and where $0 \le n < N - 1$.
The Hankel decomposition yields the complex time series $a_\ell(t)$ and $b_\ell(t)$ of in- and outgoing waves within the annulus, by making use of the orthogonality of Hankel functions:

\begin{align}
a_\ell(t) &= \frac{\pi L}{2\varDelta r}\int_{r_\text{inner}}^{r_\text{outer}}\lambda_N(r)V(r, t)H^{(2)}(Lr)rdr{ ,}\label{eq:a}\\
L &= \sqrt{\ell(\ell+1)}\text{\hspace{.33cm}and\hspace{.33cm}}
\varDelta r = r_\text{inner} - r_\text{outer}\nonumber{ ,}
\end{align}

\noindent whereby $H^{(2)}(Lr)$ is the Hankel function of second kind and $r$ is the radial coordinate in the annulus. For calculating $b_\ell(\nu)$ Equation \ref{eq:a} can be used with the Hankel function of first kind $H^{(1)}(Lr)$.
From the complex amplitudes $a_\ell(t)$, $b_\ell(t)$ we estimate the power spectra $P^\text{in}_\ell(\nu)$ and $P^\text{out}_\ell(\nu)$
%, and the total power $P^\text{tot}_\ell(\nu)$ as follows:

\begin{align}
P_\ell(\nu) = \mathcal{W}_S\left( f_\ell( t) \right)\mathnormal{ ,}
\label{eq:1}
\end{align}

\noindent where $\mathcal{W}_S(f)$ denotes the estimation process via Welch's method \citep{1967book:IEEE.....7073}, using again a Hann-window (Eq. \ref{eq:w}) with $f_\ell( t)$ representing either $a_\ell( t)$ or $b_\ell( t)$. 
As is usual for Welch's method, the time-series is further divided into $S$ equally long segments. 
A window function is applied to each segment, designed in such a way that it overlaps with half of the following segment. 
This overlap is preferred to avoid bias introduced due to the window function almost eliminating velocity signal close to the edge of individual segments.
This techniques results in $S$ individual periodograms, which are then averaged to reduce noise (and subsequently yield the estimated power spectrum).
Here we set $S=3$, as this appears to be the best trade-off between sufficient frequency resolution and noise reduction in the resulting spectrum.
The spectra $P^\text{in}_\ell(\nu)$, $P^\text{out}_\ell(\nu)$ are obtained for $f_\ell(t)=a_\ell(t)$, $b_\ell(t)$ respectively. 
Subsequently, we calculate the absorption coefficient \citep{1987ApJ...319L..27B}:

\begin{align}
\alpha_\ell(\nu) = 1-P_\ell^\text{out}(\nu)/P_\ell^\text{in}(\nu)\mathnormal{ .}
\label{eq:alpha}
\end{align}

Finally, the total power is estimated via
\begin{align}
	P^\text{tot}_\ell(\nu) = P^\text{in}_\ell(\nu)+P^\text{out}_\ell(\nu){ .}
	\label{eq:ptot}
\end{align}
	
\begin{figure}[htb!]
	\centering
	\includegraphics[width=1.\textwidth]{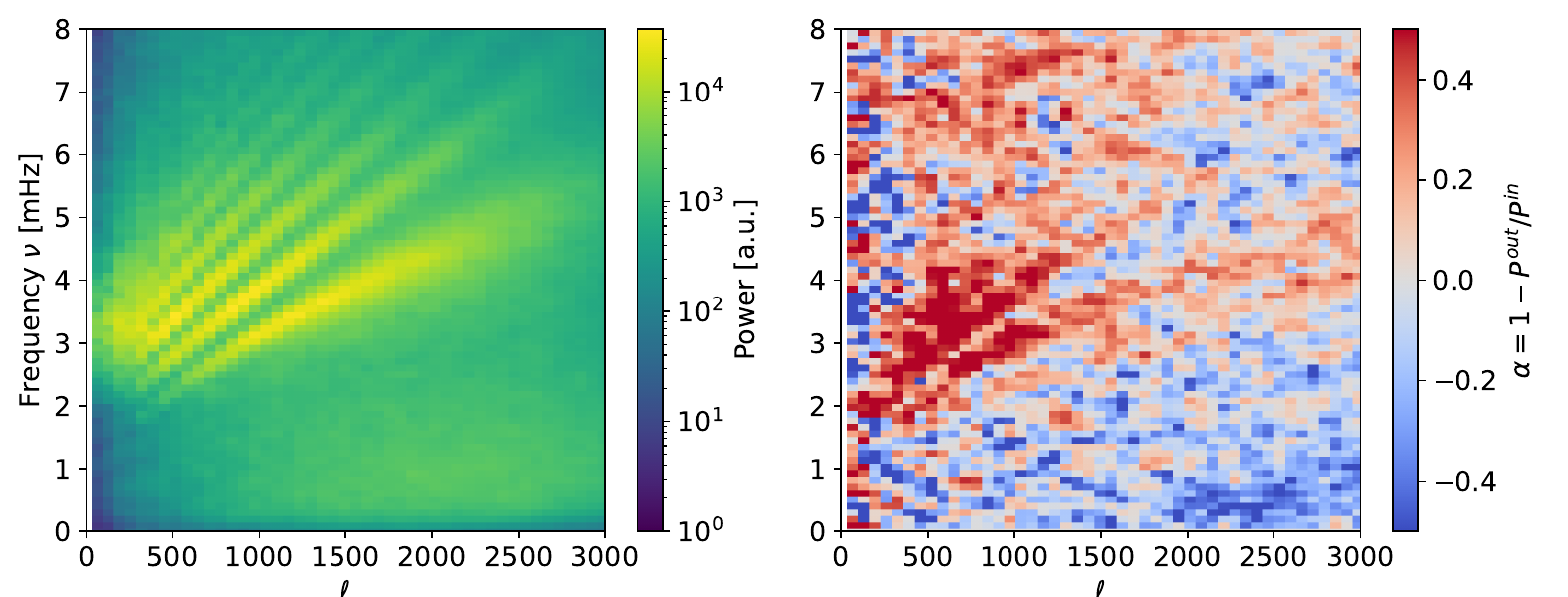}
	\caption{
		Power spectrum $P^\text{tot}$ (left) and absorption coefficient $\alpha$ (right) as a function of harmonic degree $\ell$ and temporal frequency $\nu$ for a duration of 8 hours (shown for AR 11158). 
		As can be seen, the f-mode ridge is well resolved, granulation noise is almost eliminated (due to the running difference) and other background contributions are minimal. 
		In the absorption map, red means power absorption, while blue means power emission. 
		This section of the full time series corresponds to a time after the active region emerged, which explains why absorption along ridges can be seen.
	}
	\label{fig:3}
\end{figure}

\begin{figure}[htb!]
	\centering
	\includegraphics[width=1.\textwidth]{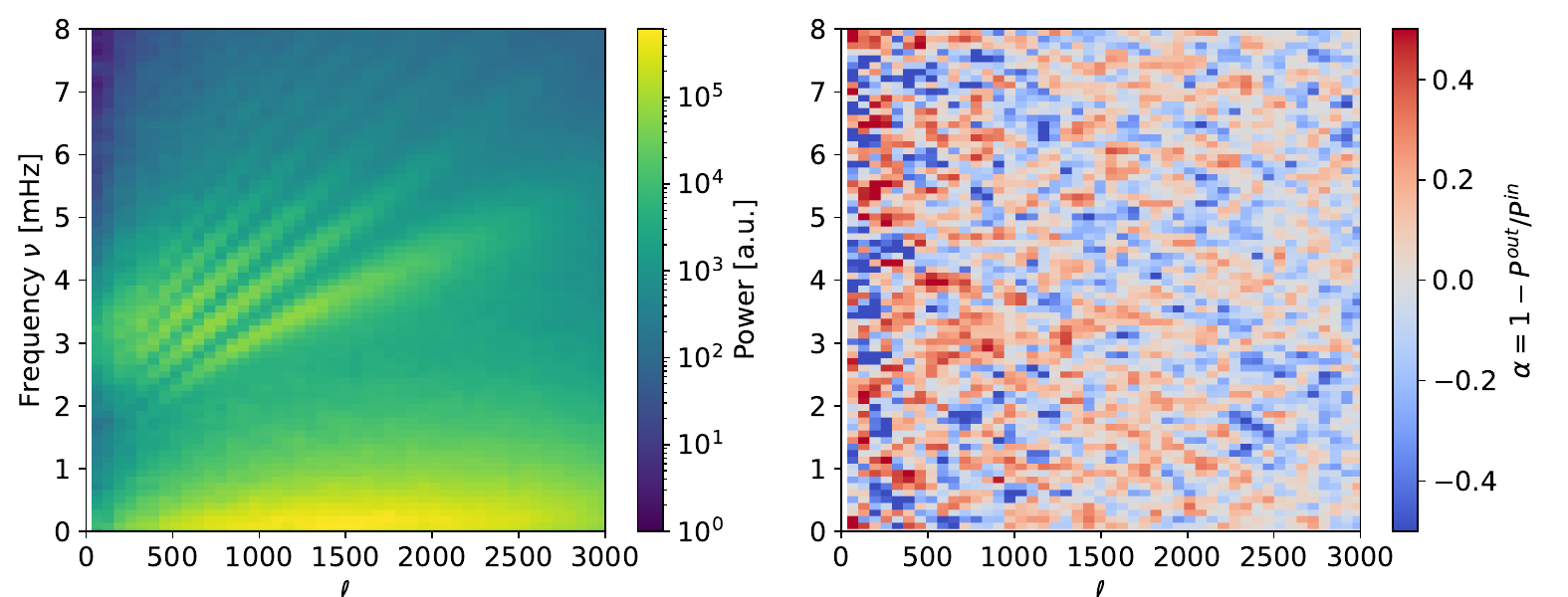}
	\caption{
		Same as Figure \ref{fig:3} but at \textbf{$t-t_\text{AR}=-24\,$h} prior to the AR emergence.
	}
	\label{fig:3.5}
\end{figure}
	
Both $P^\text{tot}_\ell(\nu)$ and $\alpha_\ell(\nu)$ are shown in Figure \ref{fig:3}.
In the left panel we then see the usual $\ell$-$\nu$-diagram with multiple ridges. 
As expected, we see barely any low frequency granulation noise, due to the running difference applied initially. 
The seemingly low resolution stems from the rather broad grid spacing of $\varDelta \ell = 59.50$ and the temporal duration of just 8 hours (the total duration is 5 days which is divided into 8 segments as mentioned). 
Furthermore, the f-mode "fans out" towards high $\ell$ \citep{2014ApJ...795L...8S}, meaning the line-width increases.
This has to be accounted for in the $\ell \in [1200, 2000]$ range, by employing an adequate fit in the following analysis. 
Comparing the total power to the absorption spectrum (right panel), we also see ridge-like structures. 
This is due to the active region's emergence towards the end of the observation, which will lead to absorption of power \citep{1987ApJ...319L..27B, 1997ApJ...486L..67C}. 
On the contrary, Figure \ref{fig:3.5} shows the absorption spectrum at $t-t_\text{AR}=-24\,$h, from which we can see indications of power absorption forming along the ridges. 
Such power absorption signatures roughly correlate with weak magnetic signals visible in the corresponding magnetogram, although no pores or other features are visible in the continuum intensity yet.
Another feature is a wide area of weak emission at very low frequencies $\nu < 2$ mHz below the f-mode frequencies and rather large $\ell$ , seemingly within the range of solar granulation (see for example \citealt{2013SoPh..282...15C, 2021ApJ...913..108W}) and propagating gravity waves. 
In this particular frequency range, the absorption coefficient $\alpha$ is rather unreliable, since not a lot of power is present in the first place. 
Either way, the region is well separated (in the frequency domain) from the f-mode ridge, ensuring that such granulation noise does not affect further analysis.
	
Error estimates are calculated as described in \cite{1981book...priestley:errors}. 
The variance of a periodogram $P$ is given as $\text{var}(P) = 4P^2$ and does not depend on the sample size $N$ (which is a general property of a $\chi_2^2$-distribution). 
Applying window functions $\lambda(t)$ to the data however, reduces the variance as follows

\begin{align}
\text{var}(P) = 4P^2\frac{\sum_{t=0}^{N}\lambda_N^2(t)}{\sum_{t=0}^{N}\lambda_N(t)}
\text{ .}
\end{align}

The apodization process also makes the $\chi^2_2$-distributed periodogram converge towards a Gaussian distribution (enabling the estimation of an actual errorbar). 
Since we use Hann windows, the variance reduces to $\text{var}(P) = P^2\cdot3/4$, with the corresponding error $\sigma_P / P = 0.866$. 
This is further reduced by averaging over $S=3$ segments (see Equation \ref{eq:1}). Furthermore, performing weighted sums over $\ell$ and $\nu$ as will be shown in the next Section finally yields $\sigma_P / P = 0.133$, or for the absorption $\alpha$

\begin{align}
\sigma_\alpha / \beta &= 0.188\text{\quad with}\\
\beta &= P^\text{out} / P^\text{in}
\text{ .}
\end{align}

This process results in a reliable estimation of errorbars as the spectra $P$ follow
approximately Gaussian distributions. Note that, in terms of terminology, the periodogram is defined as the unsmoothed square modulus of the Fourier transform of the time series. When sufficiently smoothed or averaged, it is called the power spectrum.

\section{Results}
\label{sec:4}

%To calculate the absorption coefficient $\alpha$ along the f-mode more precisely, we filtered frequencies around $\nu = \sqrt{gL/R_\odot} /2\pi$ (corresponding to the f-mode, where $g$ is the solar surface gravity, $L = \sqrt{\ell(\ell+1)}$ and $R_\odot$ is the solar radius) using a Gaussian filter with a width of $\sigma_\nu = $ 0.3 mHz (basically the area within the dashed black lines in Figure \ref{fig:3}). 
%We then averaged the filtered power over $\nu$, yielding the average f-mode power. Doing this for $P^\text{in}_\ell$ and $P^\text{out}_\ell$ separately, we can calculate $\alpha_\ell$ corresponding to f-mode frequencies only. 
%In the following, this will be done for $\ell \in [1200, 2000]$. 

\begin{figure}[htb!]
	\centering
	\includegraphics[width=1.\textwidth]{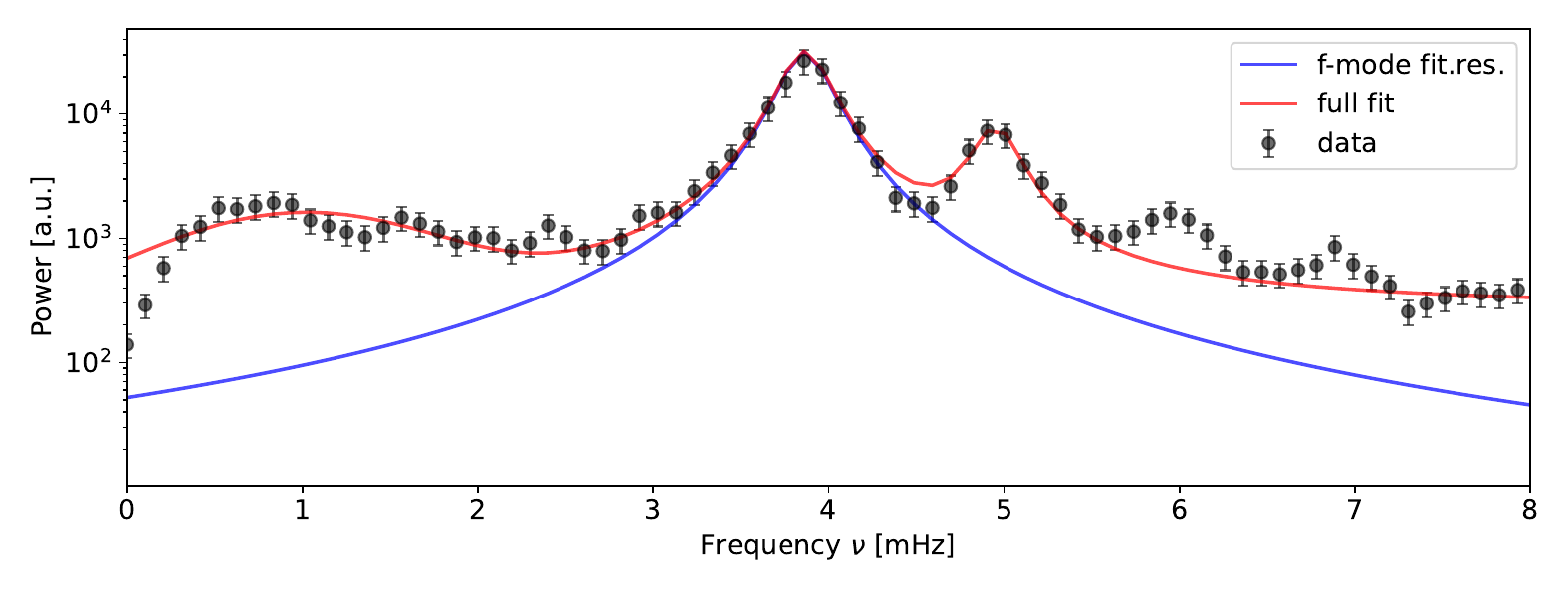}
	\caption{
		Result of fitting $P^\text{in}_\ell(\nu)$ for $\ell=1556$ for an arbitrary 8 hour segment in logarithmic scale. Black dots represent the data. The red line shows the fit result, using $M(\nu)$ as model function (Eq. \ref{eq:M}), whereas the blue line only shows the according result for $L_\text{f}(\nu) + c$, i.e. the true f-mode power spectrum plus a static background $c$. 
	}
	\label{fig:fit_res}
\end{figure}

\begin{figure}[htb!]
	\centering
	\includegraphics[width=1.\textwidth]{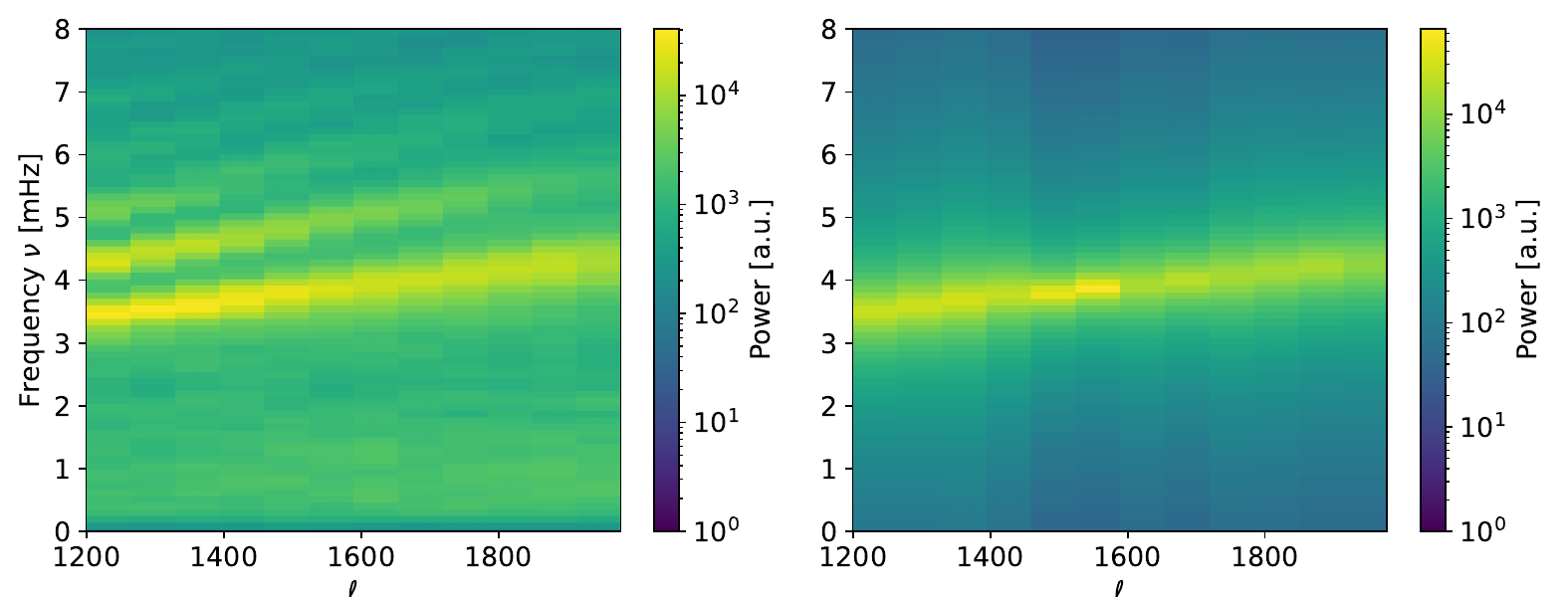}
	\caption{
		Result of fitting the spectrum around the f-mode, but for all $\ell=[1200, 2000]$ (therefore the same as Figure \ref{fig:fit_res}, but for all  $\ell=[1200, 2000]$) for an arbitrary 8 hour segment. Left shows the data $P^\text{in}_\ell(\nu)$ and right shows the according model attempt using $L_\text{f}(\nu)$ to describe the f-mode power spectrum.
	}
	\label{fig:fit_res_array}
\end{figure}

To calculate the absorption coefficient $\alpha$ along the f-mode more precisely, we first have to find solid estimates for $P^\text{in}_\ell(\nu)$ and  $P^\text{out}_\ell(\nu)$. 
For this purpose, we employ a procedure equivalent to that presented in \cite{2016ApJ...832..120S}.
Thus, we fit the entire spectrum to find the power distribution of the f-mode, using a combination of Lorentz-curves (to describe oscillatory processes within the spectrum) and Gaussian-curves plus constant (to describe background power).
Most of the power $P_\ell(\nu)$ within $\ell \in [1200, 2000]$ is made up of contributions from the f-mode ridge, the p$_1$-ridge and background, such that the most simple model $M(\nu)$ is given by two Lorentz-curves $L_\text{f}(\nu) + L_\text{p, 1}(\nu)$ plus a single Gaussian $G(\nu)$ plus a constant background $c$:
\begin{align}
	M(\nu) = L_\text{f}(\nu) + L_\text{p, 1}(\nu) + G(\nu) + c
	\label{eq:M}
\end{align}
Although there are notable contributions from higher order p-modes, they barely contribute to the f-mode power distribution itself, such that, for the sake of fit robustness, we do not consider higher orders for our fit function. 
Lastly, the frequency interval is chosen to be $\nu\in[0,8]\,$mHz, which roughly places the f-mode peak at its center, therefore ensuring an approximately symmetric line distribution.
An example of a fit result for a single value of $\ell$ (here $\ell = 1556$, exemplary for $P^\text{in}(\nu)$) is shown in Figure \ref{fig:fit_res}. 
In Figure \ref{fig:fit_res_array} we show a comparison of the available power data (also exemplary for $P^\text{in}(\nu)$) with the fit results for all spectra $\ell \in [1200, 2000]$.

For the following analysis, only the term $L_f(\nu)$ from model $M(\nu)$ will be used, since we focus on f-mode power alone while the other terms appearing in the model served the purpose of stabilizing the fit algorithm and therefore should not be considered further.
Thus, by making use of the true f-mode spectrum $L_\text{f}(\nu)$ we can estimate the energy $E_\text{f}$ within the f-mode as follows:
\begin{align}
	E_f(\nu) \propto \sum_{k}k P^\text{tot}(\nu){ ,}
	\label{eq:E}
\end{align}
therefore representing essentially a weighted sum over $k = k_y = \ell / R_\odot$. 
Still, $E_f(\nu)$ is explicitly given as a function of time $E_f(\nu)=E_f(\nu, \langle t \rangle)$, since we divided the full 6 day time series into $\langle t \rangle=8$ hour segments. 
Thus, for a smoother result, we sum over the entire spectrum $\nu$:
\begin{align}
	E_f(\langle t \rangle) = \sum_{\nu}E_f(\nu, \langle t \rangle)\text{ .}
	\label{eq:E_avg}
\end{align}
It is important that we recognize that $E_f(\langle t \rangle)$ is not only an explicit function of time, but also an implicit function of longitude, as the tracked region wanders across the solar disk as time passes, as well as a function of the $B_0$-angle showing variations of annular period. 
This fact will lead to a systematic weakening of oscillatory power when $E_f(\langle t \rangle)$ is measured close to the limb, while the $B_0$-dependence introduces an asymmetry in power between the active region and the quiet-Sun control region.
%Solar oscillations are seen in the vertical velocity component, while contributions to horizontal velocities generally stem from granulation. 
%Finally, this causes $E_f(\langle t \rangle)$ to show strong variations along the longitude, which can be accounted for by using an empirical model $\zeta$ (see \cite{2016ApJ...832..120S}), with
Here, we eliminate both of these systematic effects by using an empirical model $\zeta$ for the expected power variation across the solar disk (see \cite{2016ApJ...832..120S}):

\begin{align}
	\label{eq:zeta}
	\zeta(\cos\gamma) &= \cos\gamma\left( q + (1-q)\cos\gamma \right),\\
	\cos\gamma &= \cos\theta\cos\phi\text{ ,}\nonumber
\end{align}
in which $\theta$ and $\phi$ represent the latitude and longitude of the tracked region. 
Following the best-fit procedure carried out in \cite{2016ApJ...832..120S}, we further set $q=0.5$.
Now we can correct $E_f(\langle t \rangle)$ using
\begin{align}
	E_f(\langle t \rangle) \longrightarrow E_f(\langle t \rangle) / \zeta\text{ .}
	\label{eq:E_f}
\end{align}

\begin{figure}[htb!]
	\centering
	\includegraphics[width=1\textwidth]{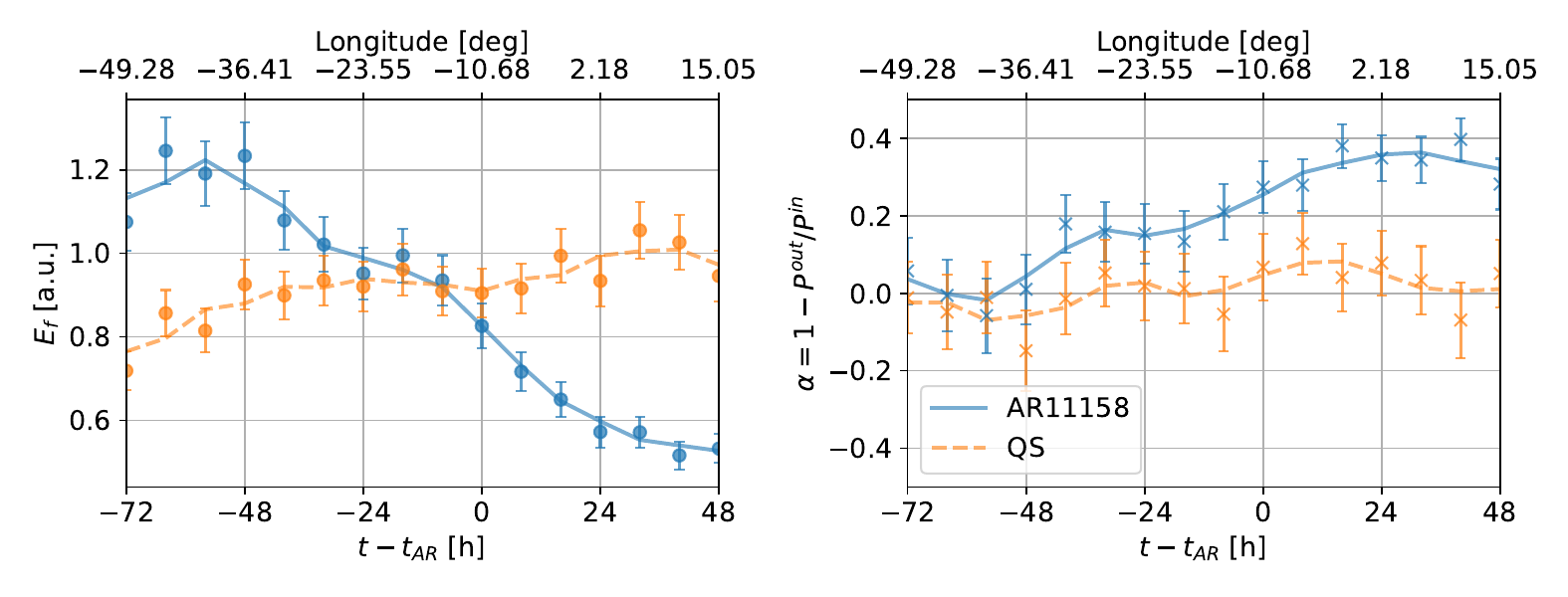}
	\caption{
		Shown are f-mode energy $E_f(\langle t \rangle)$ (left) and the absorption coefficient $\alpha$ (right) for AR 11158 and for all time segments $\langle t \rangle$ minus the emergence time of the active region $t_\text{AR}$ (namely $t - t_\text{AR}$).
		The blue curve represents data taken from the active region sample, the orange (dashed) curves shows the corresponding quiet-Sun data. 
		Curves are 3-point box-car smoothed, to reveal trends better. 
		The top axis (for both panels) shows the Stonyhurst longitude of the tracked regions.% and $t_\text{CM}$ denotes the crossing of the central Meridian (i.e. longitude $ = 0$). 
%		Bottom panels show total power of quiet-Sun regions (orange) and estimations derived from the empirical model $\zeta$ (see Eq. \ref{eq:zeta}) as a function of longitude.
%		The bottom right panel shows the variation of data values minus $\zeta$ around its expectation value (zero). 
%		The visualization of this variation serves as comparison between estimated errorbars and data variance.
	}
	\label{fig:4}
\end{figure}

The main focus of this work is to calculate power absorption as a function of time $\alpha(\langle t \rangle)$, and estimating the energy $E_f(\langle t \rangle)$ alongside $\alpha(\langle t \rangle)$ which is important to allow for direct comparison between the two and shed additional light to the findings of \cite{2016ApJ...832..120S}. 
For the correct estimation of $\alpha(\langle t \rangle)$, we employ a weighted sum over $k$ and a sum over $\nu$ for both spectra $P^\text{in}_\ell(\nu)$ and $P^\text{out}_\ell(\nu)$. 
A further correction using $\zeta$ is not required, since longitudinal dependencies are canceled out when calculating the ratio $P^\text{out}/P^\text{in}$ as in Equation \ref{eq:alpha}.
%Note that we only consider a proportional component of $E_f(\langle t \rangle)$ in Equation \ref{eq:E} which means that both energy and power essentially mean the same thing in the context of this work. 
The f-mode energy $E_f(\langle t \rangle)$ and the absorption coefficient $\alpha(\langle t \rangle)$ are both shown for AR 11158 in Figure \ref{fig:4}.

Interpreting the f-mode energy $E_f(\langle t \rangle)$ can be done by comparing both measurements for the AR and the corresponding QS region. 
Regarding the quiet-Sun energy, we can generally expect a constant, smooth distribution, providing an expectation for the 'background'-energy of the entire near side of the Sun at that time.
Therefore, any deviations of the AR distribution from its expected (QS-) distribution indicates additional physics. 
Following the conclusion of \citet{2016ApJ...832..120S}, we expect a strengthening of $E_f(\langle t \rangle)$ for the AR, about 1 to 3 days prior to its emergence (i.e. $t - t_\text{AR} = 0$).
What we observe here, is a significant power excess that can be seen before $t - t_\text{AR} < -24$ hours. %, which is similar to the observations reported in \cite{2016ApJ...832..120S}. 
For the following 4 time segments, the energy remains at nominal values before experiencing a stark reduction.
%The estimated ratio of power excess during $t = t_\text{AR}$ is $P^\text{tot}_\text{AR}/P^\text{tot}_\text{QS} = 0.871 \pm 0.074$. 
It is indeed expected that the total power in the active region is suppressed, due to the power absorption of the emerging active region itself.
%Further, the power keeps declining, down to $P^\text{tot}_\text{AR}/P^\text{tot}_\text{QS} = 0.666 \pm 0.046$ at $t - t_\text{AR} = 40$ h, which is thus also expected. 
	
Looking at the absorption coefficient $\alpha$ (right panel) for quiet-Sun (orange), we expect $\alpha\approx 0$ at all times, since no (relevant) physical process is actively absorbing or emitting power. 
%During $t - t_\text{AR} = -56$ h however, QS 11158 shows an unexpected decrease of $\alpha$ (corresponding to longitude $=-46.63^\circ$ in this case) of around $\alpha = -0.250 \pm 0.085$.
As we observe, except for minor deviations, this is indeed the case.
%Generally one needs to be cautious when analyzing Dopplergrams for longitudes less than $-40^\circ$, since the vertical velocity component is strongly reduced. 
%In fact both absorption curves (in Fig. \ref{fig:4}) follow roughly the same trend, which might mean that an overall (global) systematic effect is present during the analyzed times.
%Furthermore, all other quiet-Sun regions indicate that no negative $\alpha$ is present at that time. 
%In fact for QS 11072 we find $\alpha = 0.158 \pm 0.058$. Residue magnetic fields that are present even in quiet-Sun regions may cause increased values of $\alpha$, while negative $\alpha$ usually stem from strong sectoral asymmetries within the annulus, such as horizontal flows in, for example, only the eastern sector. 
%It is not unlikely that such unwanted effects yield imprecise estimation of $\alpha$ especially in proximity to the limb.
In principle, even weak magnetic features corresponding to, for example, short lived pores can cause momentary increases of $\alpha$, even for quiet-Sun regions. 
Concerning negative values, the only few (larger scale) mechanisms known to cause $\alpha < 0$ are acoustic glories \citep{2000SoPh..192..321D, Donea_2011} and potentially convective motions \citep{2013SoPh..282...15C}.
	
\begin{figure}[htb!]
	\centering
	\includegraphics[width=1\textwidth]{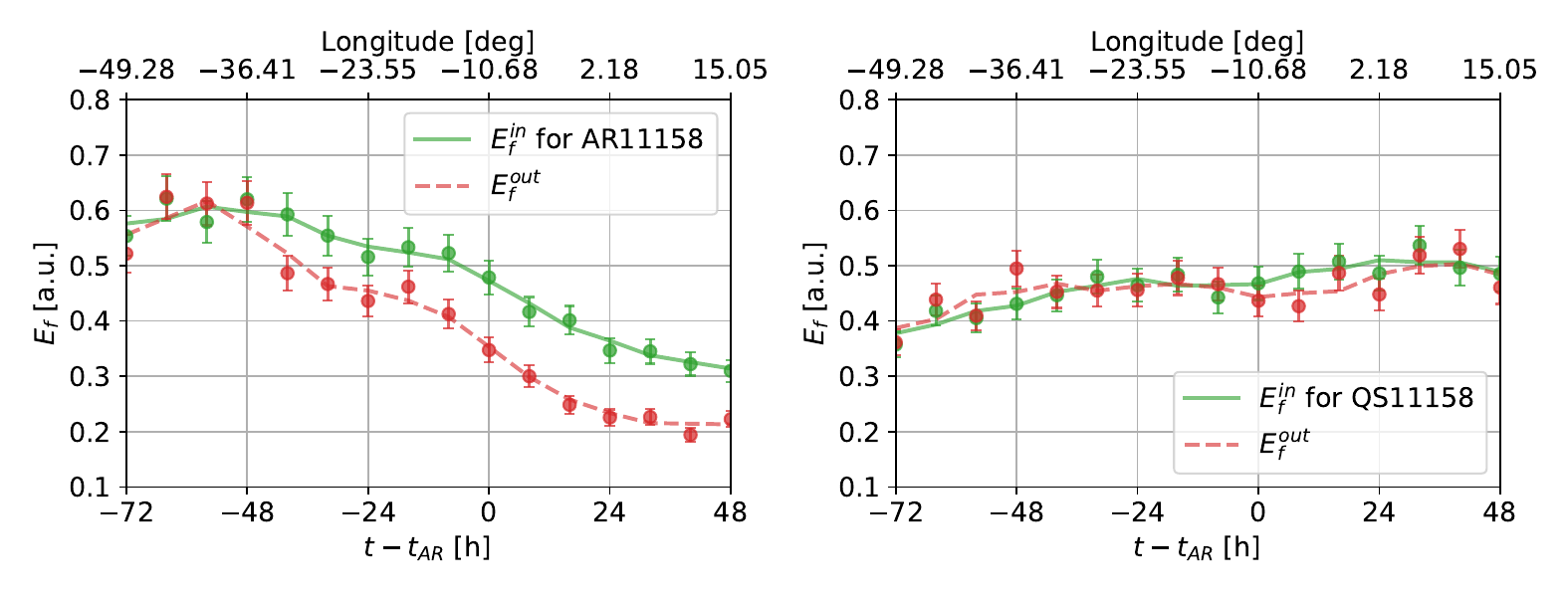}
	\centering
	\caption{
		Distributions of ingoing energy $E_f^\text{in}$ (green, solid), and outgoing energy $E_f^\text{out}$ (red, dashed) as a function of time segment minus active region emergence time $t - t_\text{AR}$. The ordinate shows units of f-mode energy, since the estimation procedure for $E_f(\langle t \rangle)$ (in Eq. \ref{eq:E_f}) was used. The left panel shows results for AR 11158, while the right panel shows results for the corresponding quiet-Sun region.
	}
	\label{fig:papb}
\end{figure}

For the active region (blue), a steady increase in absorption is observed during emergence (and arguably even before that) which is thereby in phase with the drop in $E_f(\langle t \rangle)$. 
Most notably, no power emission is detected during the strong power excess at $-72\text{ h}< t - t_\text{AR}< -40\text{ h}$. 
This is odd at first sight, although not surprising: Looking at both $E_f^\text{in}$ and $E_f^\text{out}$ individually (these are calculated via Equations \ref{eq:E_avg} and \ref{eq:E_f} using $P^\text{in}$ and $P^\text{out}$ respectively), see Figure \ref{fig:papb}, we find that $E_f^\text{out}$ is consistently weaker than $E_f^\text{in}$. 
For the quiet-Sun region we again, expect a smooth, constant distribution with $E_f^\text{in} = E_f^\text{out}$ for all times which is observed except for minor perturbations.
In order to explain the f-mode strengthening in the absence of any power emission ($\alpha < 0$), $E_f^\text{in}$ and $E_f^\text{out}$ have to increase simultaneously during the strengthening phase of $E_f(\langle t \rangle)$.
This is thus also observed as an enhancement of the energy at that time, as may be inferred by comparing the f-mode energies shown in the left and right panels of Figure \ref{fig:papb}.
Interestingly, this type of enhancement is fundamentally different from the power absorption by active regions, in which only one of the power components ($E_f^\text{out}$ in this case) is affected, while $E_f^\text{in}$ remains unchanged. 
This behavior calls for a specific investigation with the help of theoretical work (i.e. simulations), to learn the exact evolution of $E_f^\text{in}$ and $E_f^\text{out}$ during the emergence of an active region.

\begin{figure}[htb!]
	\centering
	\includegraphics[width=1\textwidth]{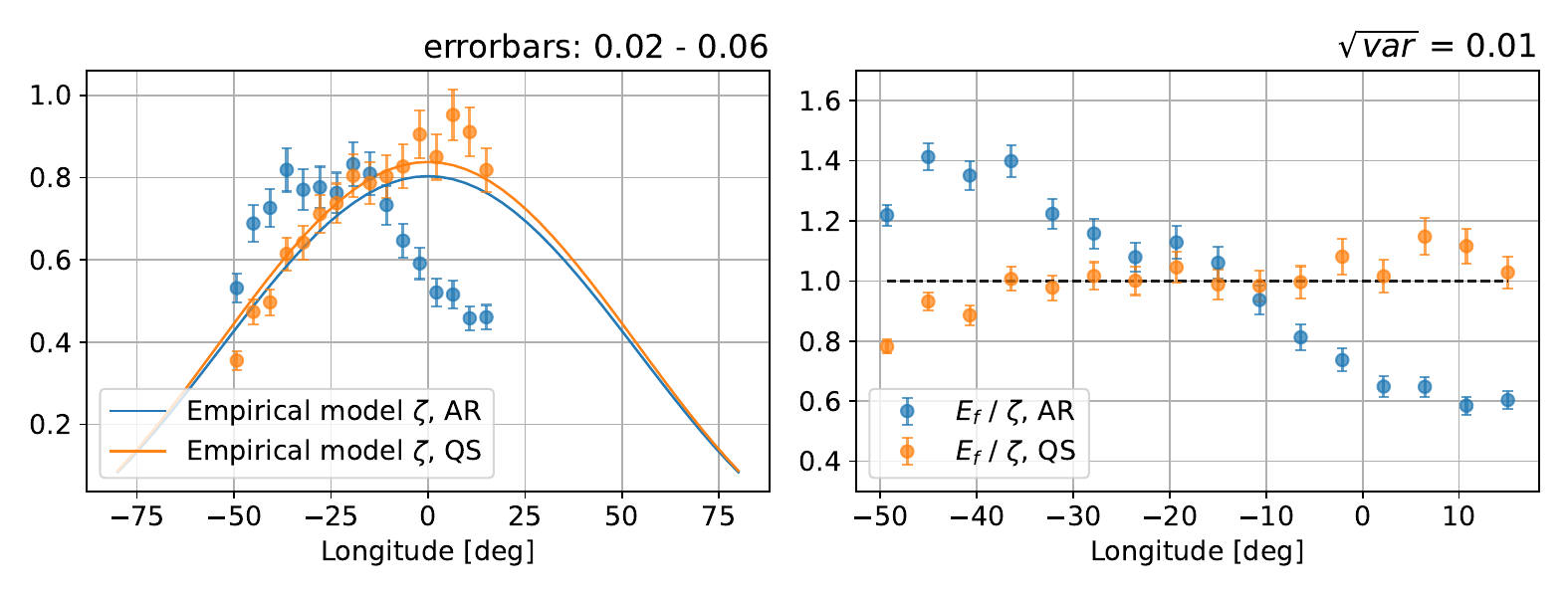}
	\centering
	\caption{
		Left: Total power $P^\text{tot} = P^\text{in} + P^\text{out}$ (dots, see Eq. \ref{eq:ptot}) and the empirical model $\zeta$ (solid lines, see Eq. \ref{eq:zeta}) as a function of longitude, shown for AR 11158 and its corresponding quiet-Sun control region. Right: Variation of $P^\text{tot}/\zeta$ around $1$ as a function of longitude. The upper right title of both panels shows the range of errors and the square root of the variance (for quiet-Sun data only) respectively.
	}
	\label{fig:var}
\end{figure}

It is important that we convince ourselves of the statistical significance of the power strengthening signal, given the fact that estimations of power spectra tend to be erratic in nature.
For this reason, we investigate both the range of errorbars and the data variance around an empirical expectation value, which is shown in Figure \ref{fig:var}. 
For an estimation of the expectation value we use the background model $\zeta$, and for the variance we use $\textrm{var} = \sum_{\langle t \rangle} (P^\text{tot}-\zeta)^2 / n_{\langle t \rangle}$ with $n_{\langle t \rangle}$ being the total amount of timesteps $\langle t \rangle$.
From comparing both, it can be seen that approximately $\sqrt{\text{var}} \approx \sigma_P$, indicating that errorbars are at the very least not underestimated. 
The systematic power asymmetry between active region and quiet-Sun introduced by $B_0\ne 0^\circ$ can be seen as a small discrepancy between the orange and blue curve in Figure \ref{fig:var}. 
As can be seen in the right panel, this discrepancy is eliminated by the correction in Equation \ref{eq:E_f}.
Finally, the total AR power (blue dots) strongly differs from its expectation value, which can be explained, as mentioned, by sunspot power absorption at longitudes close to disk center and f-mode power enhancement for lower longitudes.

%Forming the average power for $t - t_\text{AR} < -24\text{h}$ we find $P^\text{tot}_\text{AR}/P^\text{tot}_\text{QS} = 1.314\pm0.045\nonumber$, therefore observing a significant strengthening of power during that time.

\begin{figure}[htb!]
	\centering
	\includegraphics[width=1\textwidth]{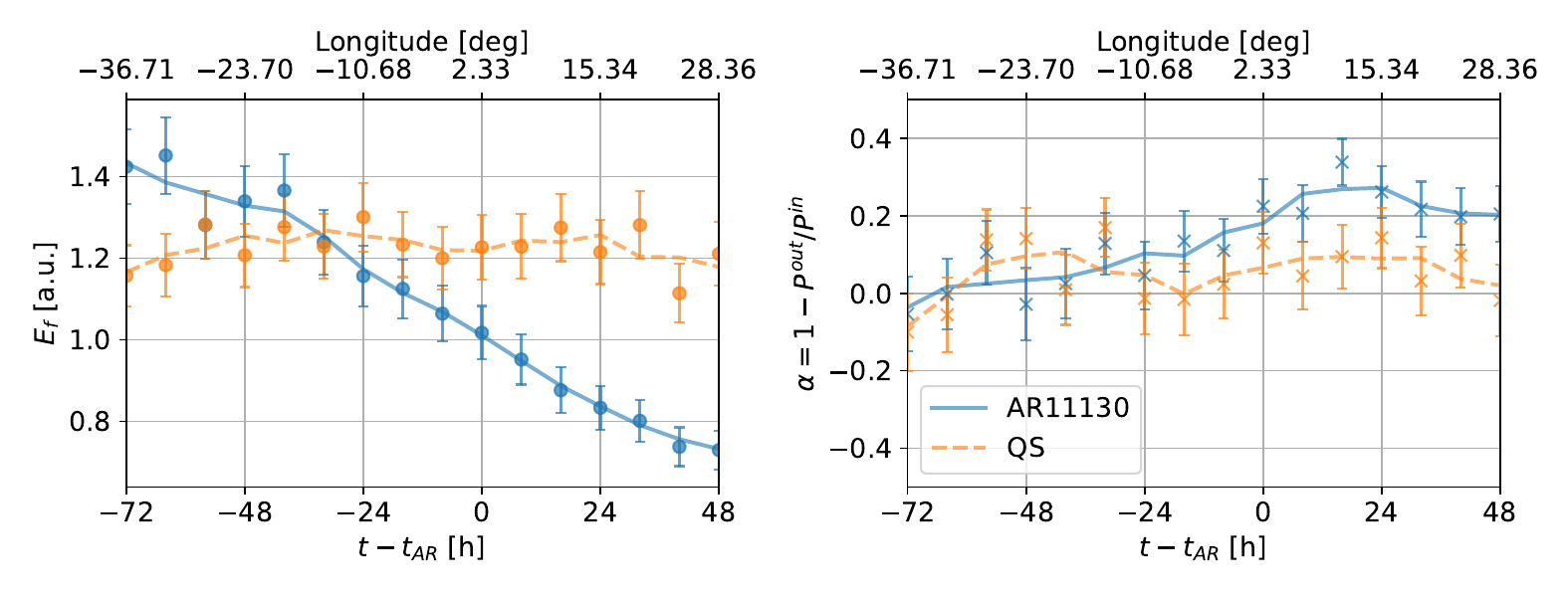}
	\centering
	\caption{
		Same as Figure \ref{fig:4}, but for AR 11130 and QS 11130.
	}
	\label{fig:5}
\end{figure}

\begin{figure}[htb!]
	\centering
	\includegraphics[width=1\textwidth]{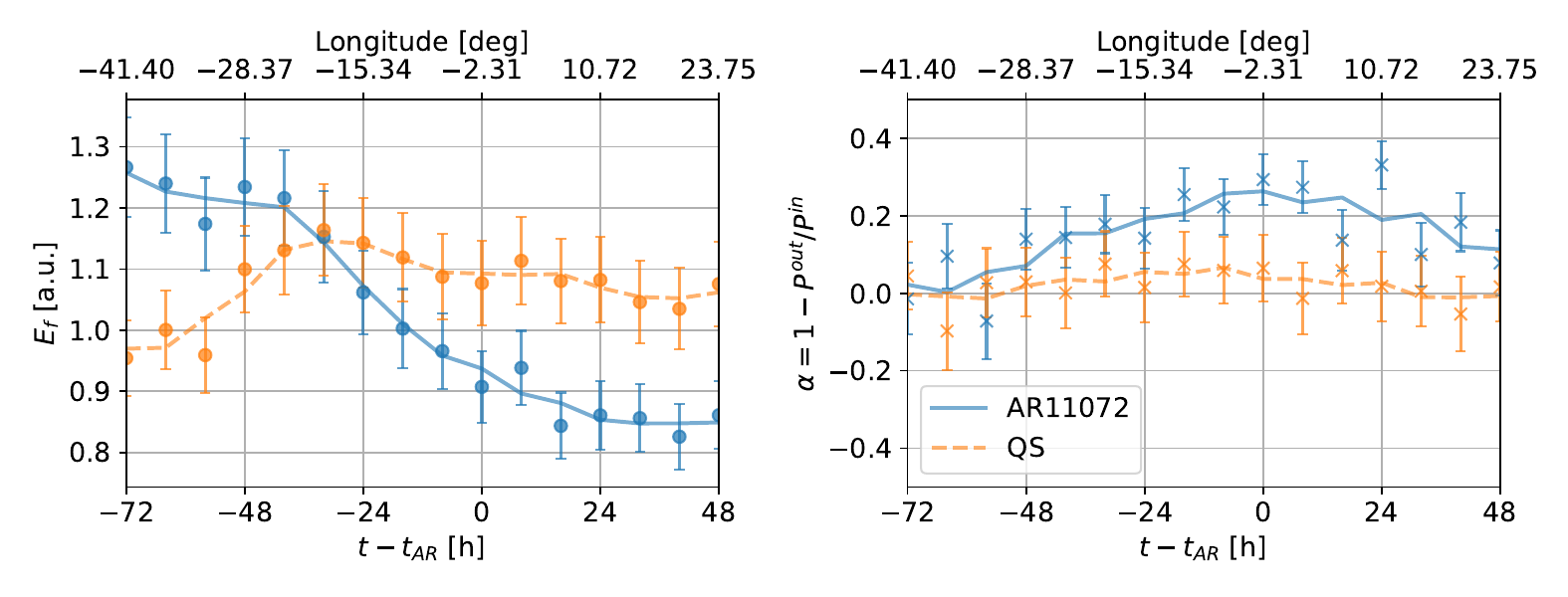}
	\centering
	\caption{
		Same as Figure \ref{fig:4}, but for AR 11072 and QS 11072.
	}
	\label{fig:6}
\end{figure}

\begin{figure}[htb!]
	\centering
	\includegraphics[width=1\textwidth]{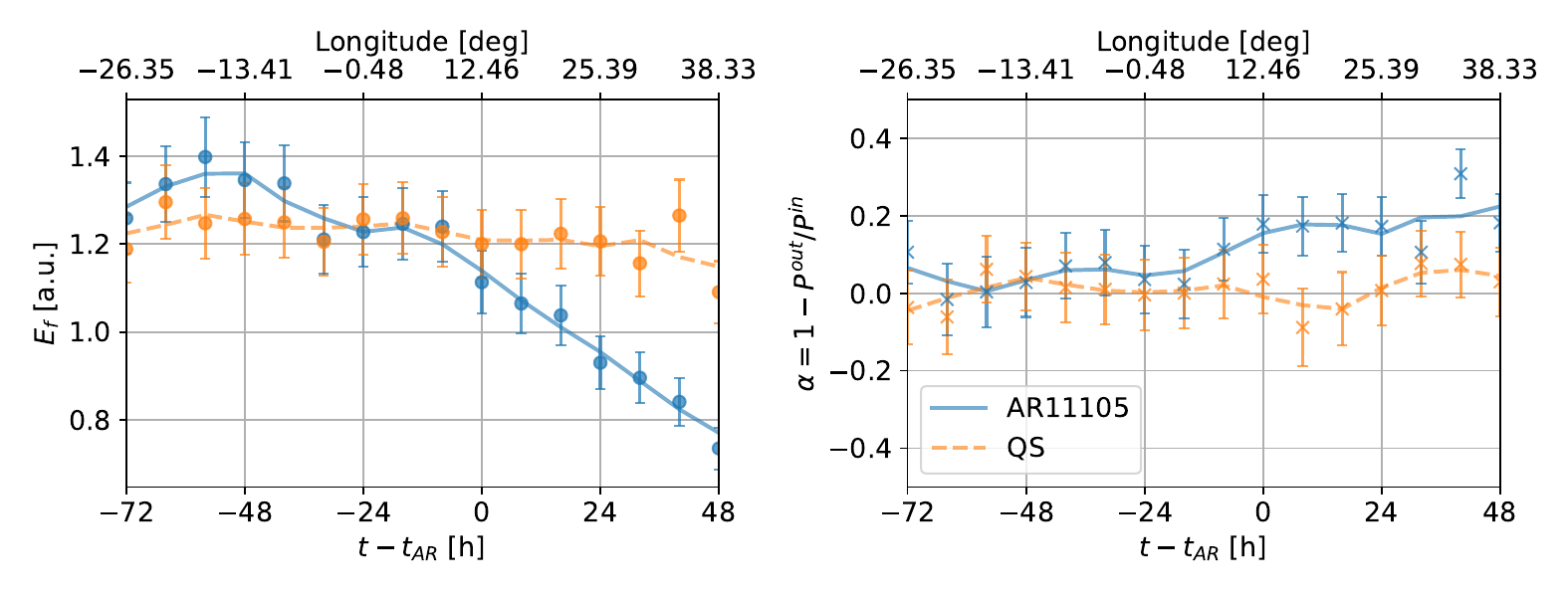}
	\centering
	\caption{
		Same as Figure \ref{fig:4}, but for AR 11105 and QS 11105.
	}
	\label{fig:7}
\end{figure}

\begin{figure}[htb!]
	\centering
	\includegraphics[width=1\textwidth]{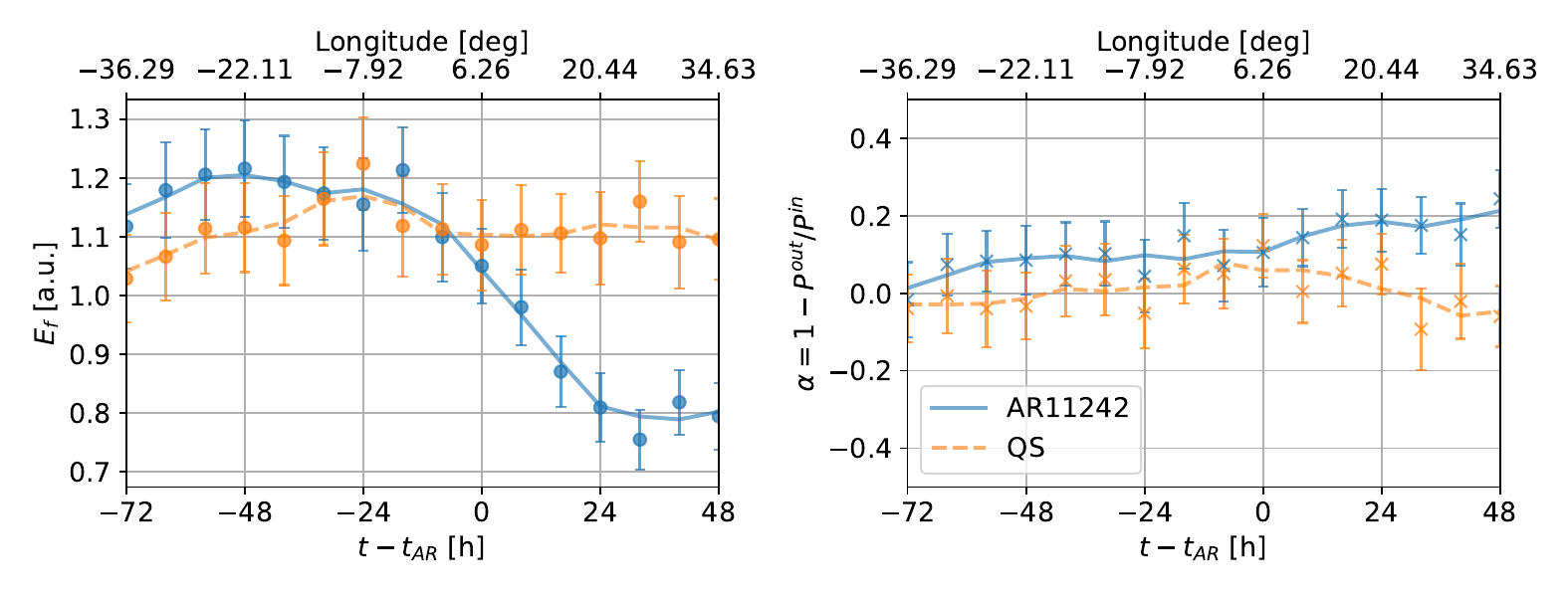}
	\centering
	\caption{
		Same as Figure \ref{fig:4}, but for AR 11242 and QS 11242. The AR and QS pair 11242 differ from the rest other ARs by their tracking rate (the Carrington-rate was used)	and the initial quiet-Sun coordinates, which was tracked two Carrington rotation prior to the emergence of AR 11242 and at the same latitude.
	}
	\label{fig:8}
\end{figure}

\begin{figure}[htb!]
	\centering
	\includegraphics[width=1\textwidth]{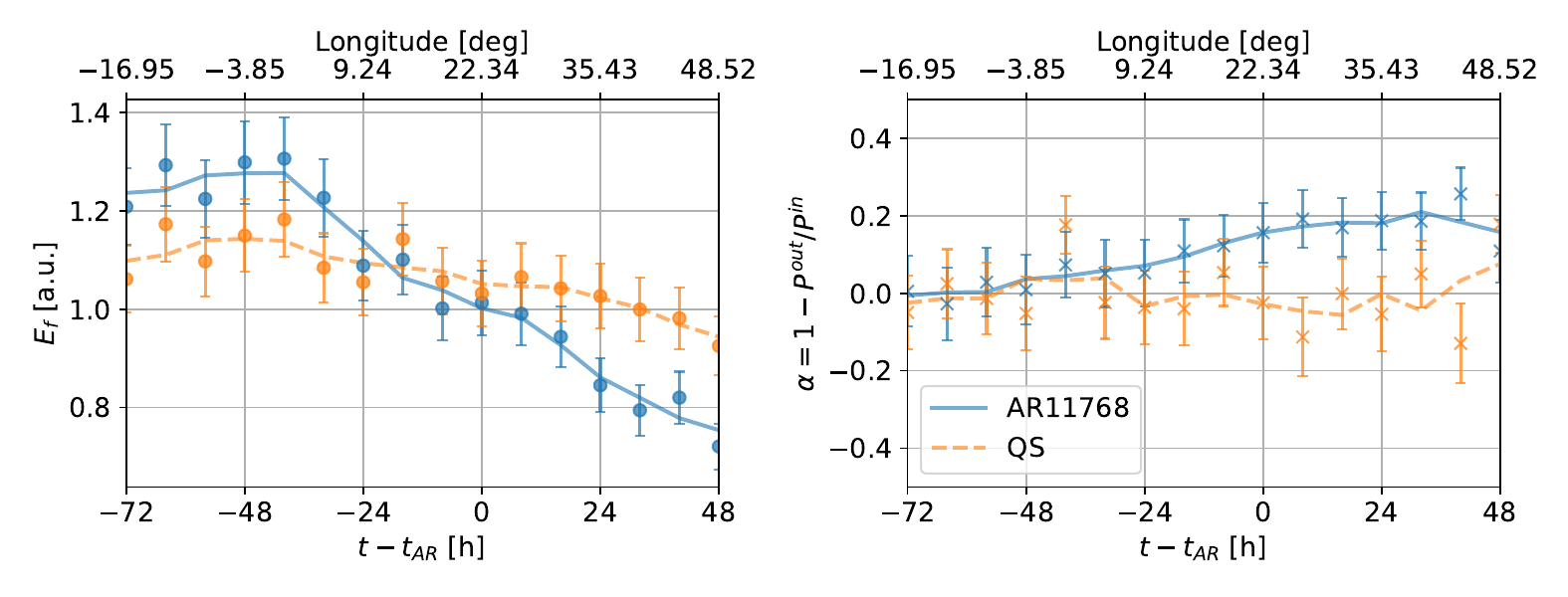}
	\centering
	\caption{
		Same as Figure \ref{fig:4}, but for AR 11768 and QS 11768.		
	}
	\label{fig:9}
\end{figure}

A set of additional active regions was tracked, according to those in \cite{2016ApJ...832..120S}, i.e. AR 11130 (Fig. \ref{fig:5}), AR 11072 (Fig. \ref{fig:6}), AR 11105 (Fig. \ref{fig:7}), AR 11242 (Fig. \ref{fig:8}) and AR 11768 (Fig. \ref{fig:9}).
We repeat the same procedure, leading up to the total power and absorption coefficient as a function of time, and show these in Figures \ref{fig:5} - \ref{fig:9}. In Table \ref{tab:2} a comprehensive overview over all results is shown.

	
We observe that the other active regions behave overall similarly to AR 11158. 
%A slight increase in $P^\text{tot}_\text{AR}/P^\text{tot}_\text{QS}$ can be seen in the average power for $t - t_\text{AR} < -24\text{h}$, where we find:
%\begin{align*}
%\text{AR 11158: } P^\text{tot}_\text{AR}/P^\text{tot}_\text{QS} &= 1.314\pm0.045\nonumber\\
%\text{AR 11130: } P^\text{tot}_\text{AR}/P^\text{tot}_\text{QS} &= 1.239\pm0.047\nonumber\\
%\text{AR 11072: } P^\text{tot}_\text{AR}/P^\text{tot}_\text{QS} &= 1.238\pm0.042\nonumber\\
%\text{AR 11105: } P^\text{tot}_\text{AR}/P^\text{tot}_\text{QS} &= 1.055\pm0.037\nonumber\\
%\text{AR 11242: } P^\text{tot}_\text{AR}/P^\text{tot}_\text{QS} &= 1.086\pm0.035\nonumber
%\end{align*}
As expected for all ARs, a depression in power at later times occurs, while $\alpha$ increases. 
While all active regions show a strengthened energy $E_f(\langle t \rangle)$ compared to their quiet-Sun counterparts, not all of them exhibit significant strengthening. %., only yielding two to three $\sigma$, above $P^\text{tot}_\text{AR}/P^\text{tot}_\text{QS} = 1$. 
Providing a slightly different tracking rate, as was done for AR 11242, as well as using a disk position offset by two Carrington rotations for the quiet-Sun control region, shows no major deviation from the behavior observed for all other regions.
The main result of $\alpha$ remaining equal to zero simultaneous to the f-mode energy strengthening is still observed however.

\begin{table}[htb!]
	\begin{tabular}{c|c|c|c}
		& AR 11158 & AR 11130 & AR 11768 \\
		\hline
		AR: $E_f(t<t_\mathrm{AR})$ & $1.16\pm0.08$ & $1.37\pm0.10$ & $1.27\pm0.09$\\
		QS: $E_f(t<t_\mathrm{AR})$ & $0.84\pm0.06$ & $1.22\pm0.09$ & $1.11\pm0.08$\\
		\hline
		AR: $\alpha(t<t_\mathrm{AR})$ & $0.04\pm0.10$ & $0.01\pm0.10$ & $0.04\pm0.10$\\
		QS: $\alpha(t<t_\mathrm{AR})$ & $-0.05\pm0.11$ & $0.03\pm0.10$ & $0.02\pm0.10$\\
		\hline
		& AR 11105 & AR 11242 &  AR 11072 \\
		\hline
		AR: $E_f(t<t_\mathrm{AR})$ & $1.34\pm0.10$ & $1.18\pm0.08$ & $1.23\pm0.09$ \\
		QS: $E_f(t<t_\mathrm{AR})$ & $1.20\pm0.09$ & $1.07\pm0.08$ & $1.03\pm0.10$ \\
		\hline
		AR: $\alpha(t<t_\mathrm{AR})$ & $0.04\pm0.10$ & $0.07\pm0.10$ & $0.06\pm0.10$\\
		QS: $\alpha(t<t_\mathrm{AR})$ & $0.00\pm0.10$ & $-0.02\pm0.10$ & $0.00\pm0.10$ \\
		\hline\hline
		& AR 11158 & AR 11130 & AR 11768 \\
		\hline		
		AR: $E_f(t>t_\mathrm{AR})$ & $0.55\pm0.04$ & $0.78\pm0.05$ & $0.80\pm0.05$\\
		QS: $E_f(t>t_\mathrm{AR})$ & $0.99\pm0.06$ & $1.20\pm0.08$ & $0.98\pm0.06$\\
		\hline
		AR: $\alpha(t>t_\mathrm{AR})$ & $0.34\pm0.06$ & $0.22\pm0.07$ & $0.18\pm0.07$\\
		QS: $\alpha(t>t_\mathrm{AR})$ & $0.02\pm0.09$ & $0.06\pm0.09$ & $0.01\pm0.09$\\
		\hline
		& AR 11105 & AR 11242 &  AR 11072 \\
		\hline 
		AR: $E_f(t>t_\mathrm{AR})$ & $0.85\pm0.06$ & $0.80\pm0.06$ & $0.85\pm0.05$ \\
		QS: $E_f(t>t_\mathrm{AR})$ & $1.18\pm0.08$ & $1.11\pm0.07$ & $1.05\pm0.07$ \\
		\hline
		AR: $\alpha(t>t_\mathrm{AR})$ & $0.19\pm0.07$ & $0.19\pm0.08$ & $0.17\pm0.08$ \\
		QS: $\alpha(t>t_\mathrm{AR})$ & $0.05\pm0.09$ & $-0.02\pm0.09$ & $0.00\pm0.09$ \\
	\end{tabular}
	\caption{
		Summary of the f-mode behavior before ($t<t_\mathrm{AR}$) and after ($t>t_\mathrm{AR}$) AR emergence. Both $E_f$ and $\alpha$ are shown for the active region as well as for the corresponding quiet-Sun region. Here, for $t>t_\mathrm{AR}$ values are averaged over a range of $t - t_\mathrm{AR}\in[-72\,\mathrm{h}, -48\,\mathrm{h}]$, selecting the time window in which f-mode strengthening is expected to occur most significantly, while for $t<t_\mathrm{AR}$, $t - t_\mathrm{AR}\in[24\,\mathrm{h}, 48\,\mathrm{h}]$ is selected.
	}
	\label{tab:2}
\end{table}

\section{Discussion and Conclusions}
\label{sec:5}

We found that the f-mode energy $E_f(\langle t \rangle)$ for AR 11158 as well as AR 11768, AR 11130, AR 11072 show a behavior that is qualitatively in line with that reported in \cite{2016ApJ...832..120S}, which is a significant enhancement during $-72 \text{h}< t - t_\text{AR} < -40\text{h}$ prior to AR emergence.
The increase in $E_f(\langle t \rangle)$ is less pronounced for other ARs, especially for AR 11105 (Fig. \ref{fig:7}) and AR 11242 (Fig. \ref{fig:8}) although still detectable.
From Table \ref{tab:2} seemingly a power enhancement is found for QS 11130 and QS 11105. However, quiet-Sun f-mode power is expected to show a long-term secular variation with the solar cycle, in such a way that the overall QS f-mode power is likely to show an anti-correlation with the magnetic cycle. QS 11130 and QS 11105 appear in isolated environments in the year 2010 during the early rising phase of the solar cycle 24. This could lead to an overall elevated power levels in these cases (see \citet{2022arXiv...2205.04419}). Long term power variations, similar to \citet{2022arXiv...2205.04419} are reflected in the results presented here by differences in the average quiet-Sun power due different points in time during the solar cycle. Signal of f-mode power strengthening, on the contrary, are characterized by a short term power increase, i.e. a departure from a flat curve (of $E_f(\langle t \rangle)$).
Still, small short term power variations can occur in quiet-Sun power. 
However, our interest is in the short-time power variability of the f-mode.
Possibly, such small variations can stem from minor deviations in the data analysis, which will be discussed below.
Investigations regarding correlations of this enhancement to the evolution of the surface magnetic flux were carried out in \cite{2016ApJ...832..120S} and yielded inconclusive results.
The current method of decomposing the observed Dopplergram is mostly analogous to that used in \cite{2016ApJ...832..120S}, but still differs in a few aspects. 
For one, using the Fourier-Hankel-decomposition to estimate spatial (total) power spectra is different from the standard spatial Fourier decomposition (into sinusoidal functions) used in ring diagram analysis.
When calculating $P^\text{tot} = P^\text{in} + P^\text{out}$, we calculate more exactly $\mathcal{W}(a) + \mathcal{W}(b)$ (see Eq. \ref{eq:1}), which represents the sum of Bessel functions of first and second kind \citep{1988ApJ...335.1015B}. 
This is mathematically different than the estimation $P^\text{tot} = \mathcal{W}(v)$ ($v$ is the velocity signal).
%This can to some extent be accounted for by calculating $P^\text{tot} = \mathcal{W}( a + b )$, resulting in a decomposition into Bessel functions of only the first kind \citep{1988ApJ...335.1015B} (thus omitting the Bessel function of second kind).
%On the contrary, using the definition $P^\text{tot} = \mathcal{W}(a) + \mathcal{W}(b)$ would include Bessel functions of first and second kind (as in the definition of Hankel functions).
However, this difference in power estimation is expected to affect mostly the power magnitude, but not the qualitative distribution.
%In this work, we used a window function to filter around $k_x = 0$ (see Fig. \ref{fig:2}), whereas originally only data at exactly $k_x = 0$ was selected. 
Additionally, the area (annulus, see Fig. \ref{fig:1}) we used is smaller, in order to retain some sensitivity of $\alpha$ to oscillations with high harmonic degree $\ell$. 
%Furthermore, \cite{2016ApJ...832..120S} used a running difference during the tracking process to eliminate granulation noise, which we disregarded here. 
%A running difference yields acceleration from the Dopplergrams, instead of velocity, which could be the reason for more pronounced power strengthening prior to AR emergence found in \cite{2016ApJ...832..120S}.
Overall, the mentioned differences in data processing methods are expected to have a minor influence on the results, as the mathematical way of decomposing the velocity signal is similar, but can still explain the subtlety of f-mode power enhancement, that we find here.
%Calculating the energy in the f-mode $E_\text{f}(t)$ as explained in \cite{2016ApJ...832..120S} seems to be better suited for finding strengthening of the f-mode. 
%However for direct comparison to $\alpha(t)$, it is instructive to use raw power $P^\text{tot}_\text{AR}$, $P^\text{tot}_\text{QS}$ and $P_\text{in}$, $P_\text{out}$, which is what is used in this study. 
Initially we mentioned that $P$ may only be approximately Gaussian distributed, such that it is unclear if $\sigma_P$ is adequately underestimated. 
Generally, there are several sources of uncertainty that may contribute to $\sigma_P$ but which are hardly quantifiable. 
Examples include instrument-conditioned uncertainties and aforementioned center-to-limb effects.
To test the accuracy of $\sigma_P$, we estimate the variance of our data $P^\text{tot}$ (exemplary for a quiet-Sun region), by using an empirical model $\zeta$ (see Eq. \ref{eq:zeta}) for its expectation value.
For QS 11158, we find values between of approximately $0.01$ for $\sqrt{\text{var}(P)}$, while for the estimated error $\sigma_P$, values range between $0.02$ -- $0.06$ (see Fig. \ref{fig:var}). 
An analysis for other QS regions yielded similar results. 
It can therefore be assumed that $\sigma_P$ is accurate, increasing our confidence in the significance of the strengthening of $E_f(\langle t \rangle)$.
Another factor to consider is the aforementioned $B_0$-angle, which is non-zero in the ARs we investigated. 
If we assume that $P^\text{tot}$ for both AR and QS follow the same trend, approximated by $\zeta$, correcting the distribution mostly eliminates this effect, except for a minor distortion in magnitude, which we expect to be much smaller than $\sigma_P$ however.
	
Regarding power absorption, it is in general less troublesome to calculate the absorption coefficient $\alpha$ than the total power $P^\text{tot}$ or energy $E$, since any result of $P_\text{out}/P_\text{in}$ automatically corrects most effects of longitudinal (or other time and location dependent) variation, projection effects (center-to-limb) and granulation noise. 
In our results (see, Figure \ref{fig:4}), it is surprising that no emission (i. e. $\alpha < 0$) accompanying the f-mode power strengthening is detected. 
On the contrary, after the emergence of the AR, its effects on the local velocity field can be seen in total power and $\alpha$ simultaneously. 
From Figures \ref{fig:5} - \ref{fig:8} we can confirm this behavior. 
In conclusion, the underlying mechanism of f-mode strengthening as observed in \cite{2016ApJ...832..120S} has thus to be different from the power absorption by sunspots as reported in \cite{1987ApJ...319L..27B} and explained in \cite{1992ApJ...391L.109S, 1993ApJ...402..721C, 2003MNRAS.346..381C}. 
We thus find with confidence that f-mode strengthening prior to AR emergence is a non-directional (at least in the sense of inward and outward propagation) phenomenon. 
An idea for a future analysis could include splitting the initial annulus into sectors, to split the power within the annulus into additional directional components (i.e. West-East, North-South). 
Lastly, from Table 2, we find weak absorption signals for some of the AR, starting to form as early as two days prior to the AR emergence. Going back to our magnetograms (Figures \ref{fig:1} and \ref{fig:1_2}), weak magnetic signatures can be seen before the AR fully emerges, and these absorption signals roughly correlate to the magnetic signatures. Therefore, an increase in $\alpha$ is not unsurprising since almost all magnetic features on the Sun exhibit power absorption (which can lead to noise in the quiet-Sun as well). This signature is not reliable however, as different ARs show an inconsistent behavior.

In conclusion, analysis of total power in the ARs 11158, 11072, 11768 and 11130 shows similar behavior to that reported in \cite{2016ApJ...832..120S}, with ARs 11105 and 11242 showing a less significant power enhancement than expected. 
Additionally, other systematic aspects of the f-mode power are found, such as an overall elevated power level for AR and QS 11130, which can be explained by long term variation of background power during the solar cycle. 
Overall, the behavior of the investigated signal can be summarized as follows: Enhancement of f-mode power at high $\ell$ one to three days prior to AR emergence, followed by depression of f-mode power that sets in just after emergence of the AR, as observed earlier.
Further studies including additional active regions are needed nevertheless, for a more statistically meaningful analysis.
	
Finding the underlying physics of these observations presented and restated here has proven to be difficult, although \cite{2020GApFD.114..196S} found from simulations that the subsurface configuration of the magnetic fields plays a major role. 
Our work serves as confirmation of the findings in \cite{2016ApJ...832..120S}, additionally yielding original results and conclusions.
In combination with the earlier work of \citet{2016ApJ...832..120S}, the current results further suggest that high spatial frequency f-mode strengthening needs to be considered as a viable precursor signal of forming ARs. 
Our results prove that the Fourier-Hankel method not only accompanies the ring diagram analysis in terms of prediction capabilities, but it also provides new insight and information, especially due to the reliability and robustness of the absorption coefficient $\alpha$ to any systematic effect.

\bibliographystyle{authordate1}
\bibliographystyle{aasjournal}

\end{article}
\end{document}